\documentclass[a4paper,twocolumn,superscriptaddress,11pt,accepted=2017-11-26]{quantumarticle}
\pdfoutput=1
\usepackage{amssymb,amsfonts,color}

\usepackage{bm} \usepackage{graphicx} \usepackage{amsmath}
\usepackage{amsthm,stmaryrd}
\usepackage{amssymb}
\usepackage{hyperref}

\begin{document}
\title{Quantum Gates Between Distant Qubits via Spin-Independent Scattering}
\date{\today}

\author{Leonardo Banchi}
\affiliation{ Department of Physics and Astronomy, University
College London, Gower St., London WC1E 6BT, UK}
\affiliation{ QOLS, Blackett Laboratory, Imperial College London, SW7 2AZ, UK}
\author{Enrico Compagno}
\affiliation{ Department of Physics and Astronomy, University
College London, Gower St., London WC1E 6BT, UK}
\author{Vladimir Korepin}
\affiliation{C. N. Yang Institute for Theoretical Physics, State University of New York at Stony Brook, NY 11794-3840, USA}
\author{Sougato Bose}
\affiliation{ Department of Physics and Astronomy, University
College London, Gower St., London WC1E 6BT, UK}


\begin{abstract}
We show how the spin independent scattering of two initially distant qubits, say, in distinct traps or in remote sites of a lattice, 
can be used to implement an entangling quantum gate between them. The scattering takes place under 1D confinement for which we consider two different
scenarios: a 1D wave-guide and a tight-binding lattice. We consider
models with contact-like interaction between two fermionic or two bosonic particles. A qubit is encoded in two distinct spins (or other internal) states of each particle.  
Our scheme enables the implementation of a gate between two qubits which are
initially too far to interact directly, and provides an alternative to photonic
mediators for the scaling of quantum computers. Fundamentally, an interesting
feature is that ``identical particles" (e.g., two atoms of the same species)
and the 1D confinement, are both necessary for the action of the gate. 
Finally, we discuss the feasibility of our scheme,
the degree of control required to initialize the wave-packets 
momenta, and show how the quality of the gate is affected by 
momentum distributions and initial distance. In a lattice, the control of quasi-momenta is
naturally provided by few local edge impurities in the lattice potential. 

\end{abstract}

\maketitle

\section{Introduction}

Recent progress in the control of the motion of neutral atoms in restricted
geometries such as traps \cite{Muldoon,Schlosser,Rydberg,Rydberg2}, 
1D optical lattices \cite{fukuhara,GreinerNew,Greif} 
and wave-guides \cite{waveguide}
has been astounding. Naturally, the question arises as to whether they can be used in a similar manner as photons are used, i.e., as ``flying qubits" for logic as well as for connecting well separated registers in quantum information processing. Quantum
logic between flying qubits exploits  their indistinguishability and assume them
to be {\em mutually non-interacting} -- hence the names ``linear optics"
\cite{KLM} and ``free electron" \cite{Beenakker} quantum computation. In fact,
for such an approach to be viable one has to engineer circumstances so that the
effect of the inter-qubit interactions can be ignored \cite{Popescu}. On the
other hand, in the context of photonic qubits, it is known that effective
interactions, engineered using atomic or other media, may enhance the efficacy
of processing information \cite{Duan-Kimble,Angelakis,Munro,Tiecke,Gorshkov}. One is thereby motivated to seek similarly efficient quantum information processing (QIP) with material flying qubits which have the advantage of {\em naturally interacting} with each other. Further motivation stems from the fact that for non-interacting mobile fermions, additional ``which-way" detection is necessary for quantum computation \cite{Beenakker} and even for generating entanglement \cite{Bose-Home}, which are not necessarily easy. Thus, if interactions do exist between flying qubits of a given species, one should aim to exploit these for QIP.

While it is known that both spin-dependent \cite{ciccarello} and
spin-independent \cite{lamata,saraga,aaaa,bbbb} scattering can entangle, it is highly
non-trivial to obtain a useful quantum gate. The amplitudes of reflection and
transmission in scattering generally depend on the internal states of the
particles involved which makes it difficult to ensure that a unitary operation
i.e., a quantum gate acts exclusively on the limited logical (e.g.
internal/spin) space that encodes the qubits. For non-identical
(one static and one mobile) particles, it has been shown that a quantum gate can
be engineered from a spin dependent scattering combined with an extra potential
\cite{guillermo}. 
An alternative approach is based on collision between matter-wave solitons that can be
used to generate entanglement between them \cite{lewenstein}. 
We will show here that one can accomplish a quantum gate
merely from the spin independent elastic scattering of two identical particles.
This crucially exploits quantum indistinguishability, as well as the equality of the incoming pair and outgoing pair of momenta in one dimension (1D). 
 In our scheme the quantum gate is only dictated by the Scattering matrix or {\em S-matrix} acting on the initial state of the two free moving qubits. This is thus an example of {\em minimal control} QIP where nothing other than the initial momenta of the qubits is controlled. Not only will it enable QIP beyond the paradigm of linear optics with material flying qubits, but also potentially connect well separated registers of static qubits. One static qubit from each register should be out-coupled to momenta states in matter wave-guides and made to scatter from each other. The resulting quantum gate will connect separated quantum registers. This may be simpler than interfacing static qubits with photons.

While quantum gates exploiting the mutual interactions of two material flying
qubits have not been considered yet in full detail, the corresponding 
situation for {\em
static} qubits has been widely studied (e.g.,
Refs.\cite{loss,jaksch,mandel,deutsch1,deutsch2,philips,grangier,Hofmann}). However,
these methods typically require a precise control of the interaction time of
the  qubits or between them and a mediating bus (e.g.,
Refs.\cite{cirac-zoller,munro-spiller,banchi}). 
Still static qubits offer the natural candidate for information storage. 
Motivated by this, and also by the high degree of control 
reached in current optical lattice experiments \cite{fukuhara,GreinerNew},  as a
second result of this paper we consider a lattice implementation  
of our gate. 
A scattering based approach for creating entanglement in lattice setups 
was considered in \cite{eprlattice}, though they assume
periodic boundaries, which are difficult to achieve, and a careful 
initialization and control of the particles' momenta. 
On the other hand, our method 
exploits a much lower control process, as  the control of quasi-momenta is
naturally provided by few local edge impurities in the lattice potential. 
This experimental proposal is particularly compelling also because
the qubit can be made either static of mobile depending on 
the tunable potential barrier on different lattice sites, thus avoiding to 
seek some mechanism to couple static and mobile particles and allowing for
both storage and computation with the same physical setup.

Our study interfaces QIP and quantum indistinguishability with two other areas,
namely the Bethe-Ansatz exact solution of many-body models \cite{Korepin} and
the 1D confinement of atoms already achieved in experiments
\cite{Tonks,chips1,chips2,Aspect,Druten-recent,Druten,article:Inguscio2013,CataniNew,Boll,Zupancic,WeitenbergFluor,GreinerNew}.

\section{Quantum gate between flying qubits}
\begin{figure}
  \includegraphics[width=.45\textwidth]{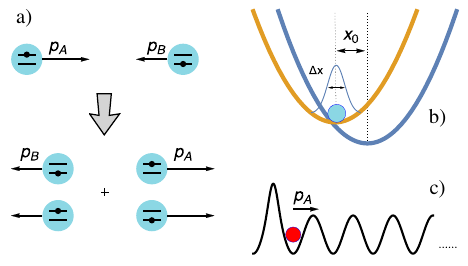}
\caption{(color online) Part 
(a) shows the nature of 1D scattering of two identical particles,
labeled as $A$ and $B$ according to their momenta directions, where some 
internal states encode the logical states of the qubit.
Incoming particles of momenta $p_A$ and $p_B$ imply outgoing particles of
exactly the same momenta. Their internal degrees of freedom on the other hand
get entangled after the collision. 
Part (b) and (c) shows two different physical implementations of the scheme
depicted in part (a). Part (b) deals with flying qubits, where
a momentum $p$ is obtained via suitably controlled local traps. 
Part (c) considers an optical lattice implementation, where a higher 
barrier on the left forces the qubit to travel to the right.
}
\label{Fig1}
\end{figure}

A two qubit entangling gate is important as it enables universal quantum computation when combined with arbitrary one qubit rotations \cite{bremner}. We consider the spin independent interaction to be a contact interaction between point-like non-relativistic particles. 
For two {\em spinless} bosons on a line (1D) the Hamiltonian 
with a delta-function interaction is \cite{Korepin}
\begin{equation}
H=-\frac{\partial^2}{\partial x_1^2}-\frac{\partial^2}{\partial x_2^2}+2 c \delta(x_1-x_2),
\label{ham}
\end{equation}
where $x_1$ and $x_2$ are the coordinates of the two particles. The above model is called the Lieb-Liniger model and has an interesting feature which we will actively exploit. This is the fact that the momenta are individually conserved during scattering. If the incoming particles have momenta $p_1$ and $p_2$, then the outgoing particles {\em also} have momenta $p_1$ and $p_2$ \cite{energy}, as shown in Fig.\ref{Fig1}(a). Thus the scattering matrix is diagonal in the basis of momenta pairs and is, in fact, only a phase which accumulates on scattering. 
The scattering matrix extracted from these wavefunctions is given,
for incident particles with momenta $p_2{>}p_1$, by \cite{Korepin}
\begin{equation}
S(p_2,p_1){=}\frac{p_2-p_1-ic}{p_2-p_1+ic}.
\label{Smat1}
\end{equation}
The phase accumulated on scattering is ${-}i\ln S(p_2,p_1)$. Note that, as
expected, for non-interacting bosons ($c{\rightarrow} 0$), their exchange causes
no phase change, while when $c{\rightarrow} \infty$
(impenetrable bosons equivalent to free fermions) 
have a ${-}1$ factor multiplying on exchange.


We consider the case of colliding particles having some 
internal degrees of freedom in which a qubit can be encoded
(Fig.\ref{Fig1}(a)).
The collision is assumed to have the form of a {\em spin independent} contact
(delta) interaction of point-like particles as in Eq.(\ref{ham}). We first
consider bosons with two relevant states $|{\uparrow}\rangle$ and
$|{\downarrow}\rangle$ of some internal degree of freedom (could be any two
spin states of a spin-1 boson, for example), and 
define the swap (permutation) operator $\Pi_{12}$ on the internal
(spin) degrees of freedom as
$\Pi_{12}(|u\rangle_1|v\rangle_2){=}|v\rangle_1|u\rangle_2$, where
$|u\rangle_1$ and $|v\rangle_2$ are arbitrary spin states of the particles 
-- so $\Pi_{12}$ is a $4{\times} 4$ matrix. 
From the swap operation we can construct the 
projectors on the symmetric (+) and antisymmetric (-) subspaces as
$\Pi_\pm = (1\pm\Pi_{12})/2$. 
For symmetric states of the internal degrees of freedom, namely eigenvectors of 
$\Pi_{12}$ with eigenvalue 1, the external degrees of
freedom also have to be symmetric and have the {\em same} scattering matrix as
spinless bosons \eqref{Smat1}. 
On the other hand, for antisymmetric spin states, the spatial
wave function of the two particles is fermionic so that the amplitude for
$x_1{=}x_2$ (the chance of a contact delta interaction) is zero implying that
they do not scatter from each other. The above observations lead to the S-matrix
$S^B(p_2,p_1)=S(p_2,p_1)\Pi_+ + \Pi_-$, namely (for $p_2{>}p_1$) 
\begin{equation}
S^B(p_2,p_1)=\frac{(p_2-p_1)-ic\Pi_{12}}{p_2-p_1+ic}.
\label{scatterboson}
\end{equation}
We also consider the case where qubit states are spin states of a spin-1/2
particle (say, electrons or fermionic atoms). This is the conventional encoding
in many quantum computation schemes. In this case the $S-$matrix was computed by
C. N. Yang \cite{Yang} to be (for $p_2{>}p_1$)
\begin{equation}
S^F(p_2,p_1)=\frac{p_2-p_1+ic\Pi_{12}}{p_2-p_1 + ic}.
\end{equation}

We consider a frame in which two qubits are moving towards each other, so that after some time they interact with the spin-independent interaction \eqref{ham}. Let us call the qubit with momentum towards the {\em right} as qubit $A$, while the qubit with momentum towards the {\em left} is called qubit $B$. 
Each qubit is in a definite momenta state, whose magnitudes are $p_A$ and $p_B$
respectively \cite{second}. Thus $p_2{=}p_A$ and $p_1{=}{-}p_B$.
The evolution of the 4 possible qubit states due to the scattering is thereby given by
\begin{align}
  S^{B/F}|{\uparrow}\rangle_A|{\uparrow}\rangle_B &=  e^{i \phi_{B/F}} |{\uparrow}\rangle_A|{\uparrow}\rangle_B      
\label{bosongate}\\\nonumber
S^{B/F}|{\downarrow}\rangle_A|{\downarrow}\rangle_B &=  e^{i \phi_{B/F}} |{\downarrow}\rangle_A|{\downarrow}\rangle_B \\\nonumber
  S^{B/F}|{\uparrow}\rangle_A|{\downarrow}\rangle_B &=  \frac{p_{AB}|{\uparrow}\rangle_A|{\downarrow}\rangle_B \mp ic |{\downarrow}\rangle_A|{\uparrow}\rangle_B}{p_{AB}+ic}
\\\nonumber
S^{B/F}|{\downarrow}\rangle_A|{\uparrow}\rangle_B &= \frac{p_{AB}|{\downarrow}\rangle_A|{\uparrow}\rangle_B\mp ic|{\uparrow}\rangle_A|{\downarrow}\rangle_B }{p_{AB}+ic}
\end{align}
where $p_{AB} {=} p_A{+}p_B$, $e^{i \phi_B}{=} \frac{p_{AB}-ic}{p_{AB}+ic}$ and $e^{i\phi_F}{=}1$.
Unless either $p_{AB}$ or $c$ vanishes, the above is manifestly an entangling
gate, as is evident from the fact that the right hand sides of the last two
lines of Eq.~(\ref{bosongate}) is an entangled state. This gate is the most
entangling (i.e., the most useful in context of quantum computation, equivalent
in usefulness to the well known Controlled NOT or CNOT gate) when $p_{AB}{\approx} c$, as then the right hand sides of the last two lines of
Eq.~(\ref{bosongate}) correspond to maximally entangled states
$\frac{e^{-i\frac{\pi}{4}}}{\sqrt{2}}(|{\uparrow}\rangle_A|{\downarrow}\rangle_B{\mp}i|{\downarrow}\rangle_A|{\uparrow}\rangle_B)$
and
$\frac{e^{-i\frac{\pi}{4}}}{\sqrt{2}}(|{\downarrow}\rangle_A|{\uparrow}\rangle_B{\mp}i|{\uparrow}\rangle_A|{\downarrow}\rangle_B)$ respectively.
 The above gates would aid universal quantum computation by means of scattering
 with both bosonic and fermionic qubits. The gates of Eqs.(\ref{bosongate}) are
 easiest to exploit as the only other requirement, namely local rotations of the
 qubit states are accomplishable by means of 
laser induced transitions between different atomic internal levels or electronic spin rotations by magnetic fields.

\subsection{Error estimates for extended packets}
The amplitudes in Eqs.(\ref{bosongate}) 
  depend only on the ratio of $p_{AB}/c$ and thereby any spread $\delta
  p_{AB}$ of the relative momenta of the incoming particles only affects the
  amplitudes as $\delta p_{AB}/c$. 
As a relevant example we consider two Gaussian wavepackets in the 
internal state
$|{\uparrow}\rangle_A|{\downarrow}\rangle_B$. Since the center of mass and relative
coordinates are decoupled in Eq.~\eqref{ham} we assume that the Gaussian packet 
can be factorized as $\tilde{\psi}(p_{\rm c.m.})\psi(p)$ where $p{=}p_2{-}p_1$ and $p_{\rm c.m.}
{=} (p_2{+}p_1)/2$. The wave-function $\tilde{\psi}(p_{\rm c.m.})$ can be ignored, as it 
provides only a global phase, while $\psi(p)$ is a Gaussian wave packet centered around 
$p_{AB}$ with variance $\sigma_p$. After the scattering, the state is 
\begin{equation}
	|\psi\rangle = \int dp \,\psi(p)\frac{p\, |{\uparrow_{A}}{\downarrow_{B}};p\rangle \mp  ic \,
	|{\downarrow_{A}}{\uparrow_{B}};p\rangle}{p+ic}~.
\end{equation}
The entanglement between the internal
degrees of freedom in the scattered state can be 
measured by the concurrence \cite{Concurrence,Amico} 
$C$. After a partial trace over the momentum, one finds that 
$C{=}\big|\int |\psi(p)|^2 \frac{2 c p}{p^2{+}c^2}\, dp\big|$,
namely $C{=}|2\Re[z]\Im[f(z)]|$, where
$z{=}\frac{c-ip_{AB}}{\sqrt{2} \sigma_p}$ and $f(z){=}\sqrt\pi e^{z^2}
{\rm erfc} (z)$. 
\begin{figure}[t]
	\centering
	\includegraphics[width=0.95\linewidth]{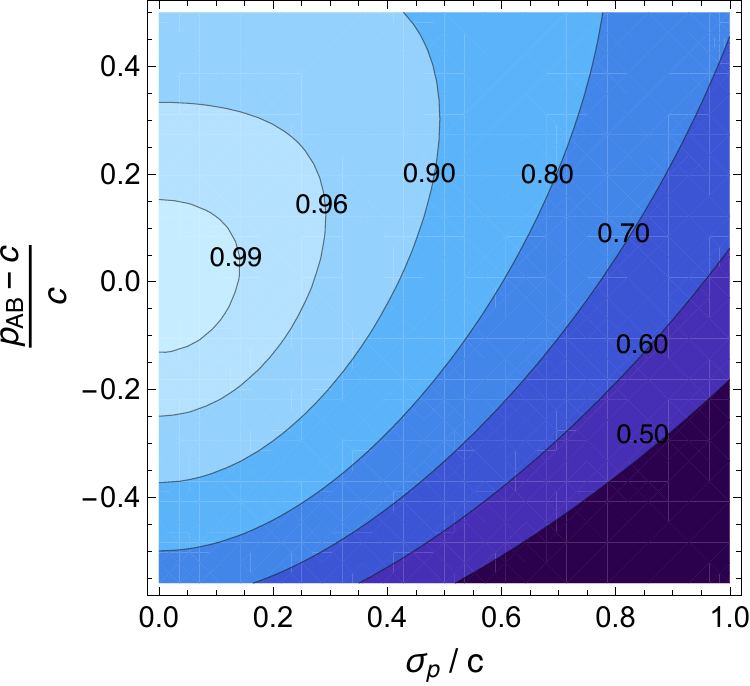}
	\caption{Concurrence between internal (spin) degrees of freedom for two Gaussian 
		wave-packets, when the relative
		momentum $p_2{-}p_1$ is peaked around $p_{AB}$ with width $\sigma_p$, and 
		$c{>}0$. Numbers indicate the value of the concurrence for those contours. 
  }
	\label{fig:fig-c}
\end{figure}
From the asymptotic expansion \cite{Abramowitz} 
$zf(z){\approx}1{-}z^{-2}/2$ one obtains that $C$ slowly decays as 
a function of $\delta{=}(p_{AB}{-}c)/c$ and $\sigma_p/c$, and that the 
case $\delta{\gtrsim}0$ is less prone to errors when $\sigma_p$ increases --
see also Fig.~\ref{fig:fig-c}.
Errors can thereby be {\em arbitrarily reduced} in principle by choosing particles with higher $c$. This is opposite to the usual paradigm of gates based on ``timed" interactions, where for a given timing error $\delta t$, stronger interactions enhance the error (while weaker interactions make gates both slower and susceptible to decoherence).

\subsection{Explicit time dependence}
In the previous section we used the scattering matrix formalism, which works in the 
asymptotic regime. Here we work out the time and space dependence more explicitly,
focusing on the bosonic case, though a similar analysis can be performed also 
for fermionic particles.
By introducing the relative $x_r{=}x_1{-}x_2$ and central $x_m{=}x_1{+}x_2$ coordinates 
we see that the Hamiltonian \eqref{ham} can be written as 
$
H=-2\frac{\partial^2}{\partial x_m^2}-2\frac{\partial^2}{\partial x_r^2}+2 c \delta(x_r)
$,
where $x_m$ and $x_r$ are decoupled. As noted in the previous section, the evolution 
in the symmetric and anti-symmetric subspaces differ only by the interaction term
-- for exclusion principle the $\delta$ interaction is effectively zero in the 
anti-symmetric space. Therefore, in these two subspaces the central coordinates have
the same evolution and can therefore be ignored. In the anti-symmetric subspace 
the relative coordinates evolve without $\delta$ interaction and therefore their 
dynamics is described \cite{TimeD} by the propagator 
$
G_t(x,y) {=} \frac{1}{\sqrt{8 \pi  i t }}\exp \left(\frac{i (x{-}y)^2}{8t }\right)
$. 
Calling $\psi_0(y_r)$ is the initial wavefunction at $t=0$, then the evolved 
wave-packet in the anti-symmetric space is 
$\psi^-_t(x_r) {=}\int G_t(x_r,y_r) \psi_0(y_r)\,dy_r$. On the other hand, in the 
symmetric subspace the particles feel the interaction and  the 
evolved wavefunction \cite{TimeD} is 
$\psi^+_t(x_r) {=} \psi_t^-(x_r) {+} \hat{\psi}_t(x_r) 
$ where $\hat{\psi}_t(x_r){=} 
\int \hat{G}_t(x_r,y_r) \psi_0(y_r)\,dy_r$, and 
\begin{align}
	\hat{G}_t(x,y) &= -\frac{c}2 \int_0^\infty  e^{-cu/2} G_t(|x|{+}|y|{+}u,0)\,du~.
	\label{Ghat}
\end{align}
If the tow particles are initially in the product state $|{\uparrow_A}{\downarrow_B}\rangle$, 
then at time $t$ they are in the state 
\begin{align}
	|\psi_t\rangle = \int dy &\,[\psi^-_t(y)+
	\frac{\hat{\psi}_t(y)}2
	|{\uparrow_{A}}{\downarrow_{B}};y\rangle \cr& + 
	\frac{\hat{\psi}_t(y)}2
	|{\downarrow_{A}}{\uparrow_{B}};y\rangle~.
\end{align}
After a partial trace on the position degrees of freedom, and using the normalization 
of the wave-function in both symmetric and anti-symmetric subspaces 
we find that the concurrence is 
$C{=} \big|\Im\big[\int dy \, \psi^-_t(y)^*\hat{\psi}_t(y)\big]\big|$.

\begin{figure}[t]
	\centering
	\includegraphics[width=0.95\linewidth]{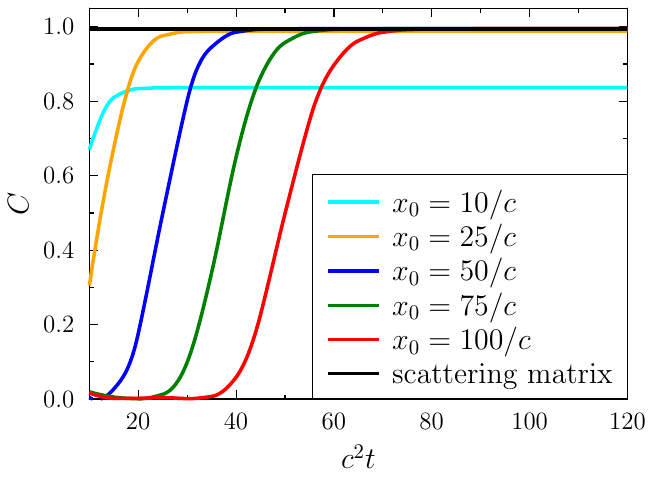}
	\caption{Concurrence as a function of time for different values of the initial position, 
		and as predicted from the scattering matrix formalism ($C{\approx}0.995$). 
		We used $\sigma_x{=}10/c$ and $p_{AB}{=}c$.
	}
	\label{fig:time}
\end{figure}

We consider now two distant Gaussian wavepackets, centered around 
positions $x_1{=}{-}x_0/2$ and $x_2{=}x_0/2$, with width $\sigma_x/\sqrt2$, 
and propagating with 
speed $v{>}0$ and ${-}v$ respectively, so that the relative momentum is $p_{AB}{=}{-}2v$. 
If $x_0{\gg}\sigma_x$, and $vt{-}x_0{\gg} \sigma_x$, then we
can assume that the particles are non interacting, both initially and at time $t$. 
With this assumption we can perform the integral in Eq.~\eqref{Ghat} analytically by 
substituting $|x|{\approx} x$ and $|y|{\approx} {-}y$. Then, the concurrence
can be calculated exactly. From the solution, we find that the explicit dependence 
on $x_0$ and $t$ disappear, without further approximations, and we get 
the same expression for $C$ as obtained from the scattering matrix formalism. 
Therefore, the predictions of the scattering matrix formalism, discussed in the 
previous section, are accurate enough irrespective of $c$, provided that the initial 
and final wave-packets are non-overlapping. This is shown explicitly in 
Fig.~\ref{fig:time} where the concurrence is evaluated numerically without 
approximations. Note that since $\sigma_p{\propto}(\sigma_x)^{-1}$ the optimal 
conditions are
$x_0{\gg}\sigma_x{\gg}c^{-1}$ and $p_{AB}{=}c$. 

\subsection{Implementation via flying qubits}
One of the most promising implementation of our gates is with neutral
 bosonic/fermionic atoms. The delta function interaction we use is, in fact,
 very realistic and realizable for neutral atoms under strong 1D confinement
 \cite{Olshanii}. $^{87}$Rb atoms have
 already been strongly confined to 1D atomic waveguides leading to delta
 interactions \cite{Tonks}. For $^{87}$Rb, with a 3D scattering length $a\,{\simeq}
 50{\AA}$ an axial (for 1D) confinement of $\omega_{\perp}{\simeq} 100$kHz gives
 (using e.g. Refs.\cite{Olshanii,Busch}) $c\,{\simeq} 10^6$ m$^{-1}$. Velocities of
 atoms in 1D waveguides ({\em c.f.} atom lasers \cite{Aspect}) can be mm
 s$^{-1}$, which translates to $p_{AB} {\simeq} 10^6$ m$^{-1}$
 (in units of wavenumbers). Thereby, $p_{AB}{\approx} c$ 
 for {\em optimal gates} is achievable in current technology \cite{comment}. 
Deviation from the 1D effective $\delta$-potential are expected when the
condition $p_{AB}{\ll}\sqrt{\hbar\mu\omega_\perp}$ is not satisfied 
(where $\mu$ is the atomic mass). In that case the scattering matrix has still the
form Eq.(\ref{bosongate}) where $c$ shows a (weak) dependence  on $p_{AB}$ \cite{Olshanii}.
The optimal gate is then found by solving $p_{AB}{=}
c(p_{AB}) {\simeq}c{-}\zeta_{3/2}(\mu\omega_\perp/\hbar)^{{-}3/2}
(c\,p_{AB}/4)^2$, where $\zeta$ is the Riemann zeta function. 
 State independent waveguides for two spin states have been met \cite{Druten-recent} in magnetic waveguides (trivially possible in optical waveguides/hollow fibers). Our gate will be an extension of collision experiments between different spin species \cite{Zwierlein} with pairs of atoms at a time. Launching exactly two atoms towards each other in 1D should be feasible 
 with microtrap arrays \cite{Muldoon,Schlosser,Marray}
 or in atom chips \cite{chips1,chips2} and is also a key assumption in many works
 \cite{Popescu,Calarco,Jaksch,Owen}. For example, 
 our gates can be made with the technique of 
 Ref.\cite{Calarco} whereby atoms are trapped initially in potential dips inside a larger well and let to roll towards each other in a harmonic potential to acquire their momenta (note that our gate scheme is completely different from Ref.\cite{Calarco}, where the atomic motion is guided by internal states). 
The initial position of the particles $x_{B}{=}{-}x_A{=}x_0$ can be tailored so
that their relative momentum has minimum variance $\Delta p_{AB}$ 
when the particles reach the collision point
$(x{=}0)$. As $\omega_z{\ll}\omega_\perp$ (where $\omega_z$ is the frequency of
the longitudinal harmonic confinement) the collision does not feel the longitudinal potential, 
so it is approximated by Eq.(\ref{bosongate}). 
As shown previously, the generated entanglement is very high ($C{\simeq}1$) provided
that $\eta{\sim}\Delta p_{AB}/p_{AB}{\sim}\Delta x_0/x_0 {\ll}1$.
A different approach (depicted is Fig.~\ref{Fig1}b) consists in suddenly 
moving the local trapping potentials so that the particles in $A$ and $B$ 
move towards their new potential minima. 
As in the previous case, 
wave-packets with well-defined and tunable momenta are 
created by switching off the potential 
when they reach the minima where their momentum uncertainties  are minimal.

\section{Quantum gates between distant stationary qubits}
A discrete variant of the system with a $\delta$-interaction is the Hubbard
Hamiltonian \cite{Book:Gaudin2014,Korepin}: 
\begin{equation}
  H {=} \sum\limits_{j,\alpha}\frac{J_j}{ 2} \left[a_{j,\alpha}^\dagger a_{j+1,\alpha} + {\rm h.c.} \right]
  + \sum\limits_{j,\alpha,\beta}\!\frac{U_j^{\alpha\beta}}{2}  n_{j,\alpha}
  n_{j,\beta}~,
  \label{e.BH}
\end{equation}
where $\alpha{=}\{\uparrow,\downarrow\}$ labels two internal degrees of freedom
of the particles. We call $N$ the length of the chain.
In a lattice setup, free-space evolution is replaced with particle hopping. 
Particle collisions lead to a scattering matrix which,
for uniform couplings $J_j{=}J$, $U^{\alpha\beta}_j{=}U^{\alpha\beta}$,
is given by Eq.\eqref{scatterboson}  
with the substitutions
\cite{Book:Gaudin2014,article:Krauth} 
\begin{align}
p_j&\rightarrow\sin
p_j,& c &\rightarrow U^{\alpha\beta}/J.
\label{subst}
\end{align}
A maximally entangling gate 
is therefore realized  when $\sin p_1{ -}\sin p_2 {\approx} 2U$, with
$U{=}U^{\uparrow\downarrow}/(2J)$. 
In particular, $p_{1}{=}\sin^{-1}U$
when $p_1{=}{-}p_2$. 

The Hubbard Hamiltonian \eqref{e.BH} naturally models cold bosonic/fermionic 
atoms in optical lattice
\cite{Book:Sanpera2012}. 
Owing to single atom addressing techniques \cite{article:Weitenberg2011}  $^{87}
Rb$ atoms in different lattice sites can be initialized in either two
distinguishable hyperfine internal states $\vert{\downarrow}\rangle{\equiv}
\vert{F{=}1,m_F{=}{-}1}\rangle$ and
$\vert{\uparrow}\rangle{\equiv}\vert{F{=}2,m_F{=}{-}2}\rangle$. 
The coupling constants $U_j^{\alpha\beta}$ depend on the strength
$g_{\alpha\beta}$ of the effective interaction between cold atoms 
\cite{Jaksch1998}. These parameters 
are usually experimentally measured
\cite{article:Egorov2013,article:Weiner1999} and can be tuned by Feshbach
resonances \cite{article:Marte2002}. Spin-exchange collisions are highly
suppressed due to the little difference (less than 5\%) between singlet and
triplet scattering length of $^{87}Rb$
\cite{Book:Sanpera2012}. The one-dimensional regime is
obtained by increasing the harmonic lattice transverse confinement
($\omega_\perp/2\pi {\simeq} 18$ kHz  see
\cite{article:Inguscio2013,article:Bloch2005} for typical values). We obtain
the 1D pseudo-potential coupling constants $g_{\alpha\beta}$ from the 3D
measured values \cite{article:Egorov2013} following \cite{Olshanii},
respectively $g_{\uparrow\uparrow}{=}1.14{\times}10^{-37}{\rm J\,m}$,
$g_{\uparrow\downarrow}{=}1.12{\times}10^{-37}{\rm J\,m}$,
$g_{\downarrow\downarrow}{=}1.09{\times}10^{-37}{\rm J\,m}$. 
The internal spin state and the position of particles are detected by
fluorescence microscopy techniques 
\cite{article:Weitenberg2011,article:Bakr2009}.
The parameters $U_j^{\alpha\beta}$ and $J_j$ can be physically controlled in
optical lattice systems locally  varying the depth of the optical potential
\cite{article:HubbardToolbox2004}. Arbitrary optical
potential landscapes are generated directly projecting a light pattern by using
holographic masks or micromirror device \cite{GreinerNew,article:Bakr2009}. 
In particular, $U^{\alpha\beta}{\simeq}
\sqrt{2\pi}\left(g_{\alpha\beta}/\lambda\right)\left(V_0/E_R\right)^{1/4}$  and
$J_j\simeq \left(4/\sqrt{\pi}\right)E_R
\left(V{/}E_R\right)^{3/4}\exp\left[-2\left(V{/}E_R\right)^{1{/}2}\right]$
where $\lambda$ is the laser wavelength, $V_0$ is the lattice depth and $E_R$ is
the recoil energy \cite{Jaksch1998}. 

For flying qubits, in Sec.~2 we considered a fixed $c$ and we tuned $p_j$ to obtain the 
desired gate. In a lattice, on the other hand, $U_{\alpha\beta}$ can 
be controlled precisely, while the creation of a wave-packet requires the
control and initialization of many-sites. 
This kind of control can be avoided 
by initially
placing two particles at the two distant boundaries 
of the lattice (particle $A$ on the left and particle $B$ on the right) 
and locally tuning the coupling
$J_0$ between the boundaries and the rest of the chain 
\cite{article:Banchi2011} (all the other couplings are uniform $J_j{=}J$). 
An optimal choice of $J_0/J$ has a twofold effect
\cite{article:Banchi2011}: 
firstly it generates two 
wave-packets whose momentum distribution is Lorentzian,  narrow around 
$p_A{=}{-}p_B{\simeq}{\pm}\pi/2$, respectively,  
and with a width dependent on $J_0$; 
secondly it generates a
quasi-dispersionless evolution, allowing an almost perfect reconstruction of 
the wave-packets after the transmission (occurring in a time ${\approx} N/J$) 
to the opposite end. In this scheme (shown in Fig.~1c), the particles start from opposite 
locations, interact close to the center of the chain and then reach 
the opposite end where the wave-function is again almost completely localized, 
allowing a proper particle addressing. 
Since $p_{A/B}$ is fixed, a high amount of entanglement is generated 
when $U{=}|\sin p_{A/B}|{=}1$. 
For ${}^{87}$Rb we found that the latter condition is satisfied, {\it e.g.}, 
when $V_0/E_R{\simeq}2.2$, giving also $J/h{\simeq}240 {\rm Hz}$.

\begin{figure}[t]
\centering
\includegraphics[width=0.9\linewidth]{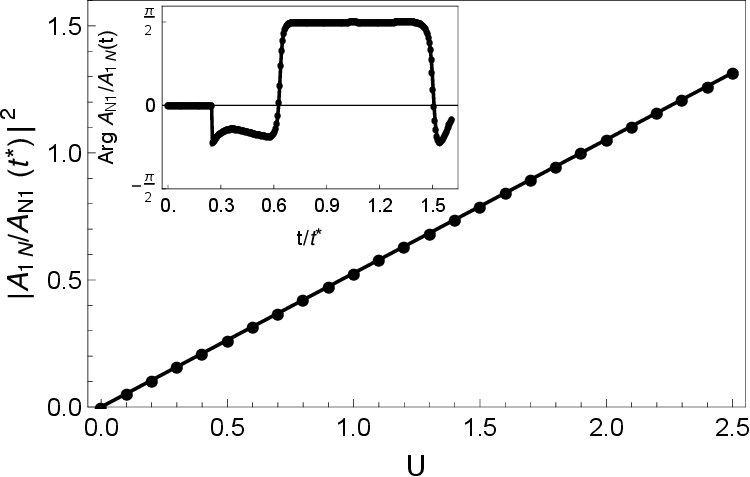}
\caption{Ratio between the probabilitites $| A_{1N}^{\uparrow\downarrow}(t)|^2$ and $| A_{N1}^{\uparrow\downarrow}(t)|^2$  as a function of the interaction parameter $U\equiv U^{\uparrow\downarrow}$ (in units of $J$), evaluated at the gate time $\tilde{t}\simeq 26.08/J$. The chain length is $N=21$ and $U$ is set to the optimal value $U=2 U^{\rm opt}=2\times 0.95$. (inset) Phase difference between the amplitude probability $A_{N1}^{\uparrow\downarrow}(t)$ and $A_{1N}^{\uparrow\downarrow}(t)$ as a function of the time $t/\tilde{t}$. }
\label{fig:AmplitudeAndPhase}
\end{figure}

\begin{figure}[t]
\centering
\includegraphics[width=0.9\linewidth]{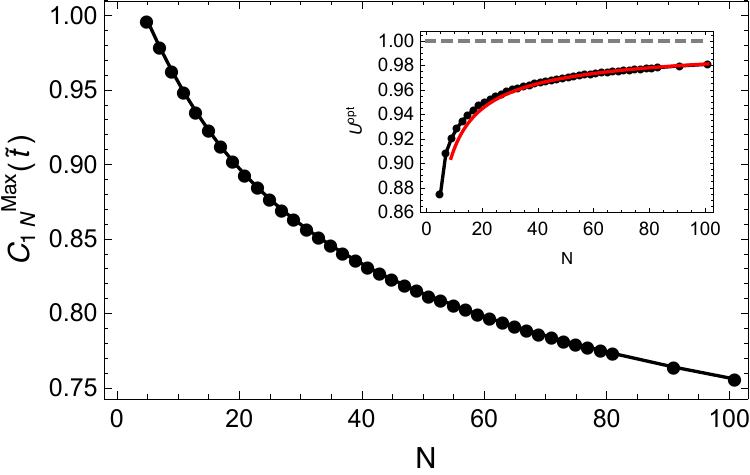}
\caption{Scaling of the maximum of concurrence as a function of the chain
length $L$. 
(inset)
Optimal inter-particle interaction strength $U^{\rm opt}$ as a function of the
chain length $N$. The numerical value is found via a linear fit over the data
of ratio $|
A_{1N}^{\uparrow\downarrow}/A_{N1}^{\uparrow\downarrow}|(\tilde{t})$ as a
function of the inter-particle interaction, $U\equiv U^{\uparrow\downarrow}/J$.
The red line is the fit function $1-0.41 N^{-2/3}$ over the data.}
\label{fig:UoptL}
\end{figure}

In this scheme there are two error sources. The first is due to 
the transmission quality, though it is above $85\%$ even for long 
chains \cite{article:Banchi2011}. 
The second one is due to the finite
width of the Lorentzian momentum profile around $|p|{=}\pi/2$ which, in turn,
yields slightly different gates for different momentum components. 
To quantify the amount of these errors 
we evaluate numerically the joint probability amplitude
$A^{\alpha\beta}_{ij}(\tilde t)$ to have
respectively particle $A$ in sites $i$ and particle $B$  in $j$ as function of 
the inter-particle interactions $U$. The indices $\alpha, \beta$ refer to
the initial internal state of particles $A$ and $B$, 
$\tilde t$ is the transfer time, and the initial condition is 
$A^{\alpha\beta}_{1N}(0){=}1$. 
As shown in Fig. \ref{fig:AmplitudeAndPhase}, we find 
$A^{\uparrow\downarrow}_{1N}{/}A^{\uparrow\downarrow}_{N1}(\tilde t){=}
{-}i U/U_{\rm opt}$ for distinguishable particles, 
where $U_{\rm opt}$ is the value of $U^{\uparrow\downarrow}$  that optimises the transformation
\eqref{bosongate} at time $\tilde{t}$. 
This optimal value is found numerically via a linear fit over the data, and
it slightly differs from the analytic 
prediction $U_{\rm opt}{=}1$ because of finite size effects. 
More precisely, in the inset of Fig. \ref{fig:UoptL} we show
that  $U_{\rm opt}$ scales with the length of the chain $N$, towards the value
$U_{\rm opt}{\to}1$, in agreement with the analytical prediction. 
For indistinguishable particles we obtain that 
$ A^{\alpha\alpha}_{11}/A^{\alpha\alpha}_{1N}(\tilde{t})$ is zero for
$\alpha{=}{\uparrow},{\downarrow}$ and for any value of 
$U_{\uparrow\downarrow}/J$. Therefore, apart from a global damping factor
due to the non-perfect wave-packet reconstruction, the resulting transformation
is in agreement with the gate \eqref{bosongate}, with the substitution
\eqref{subst} and $p_1{\simeq}\pi/2$.

\begin{figure}[t]
\centering
\includegraphics[width=0.95\linewidth]{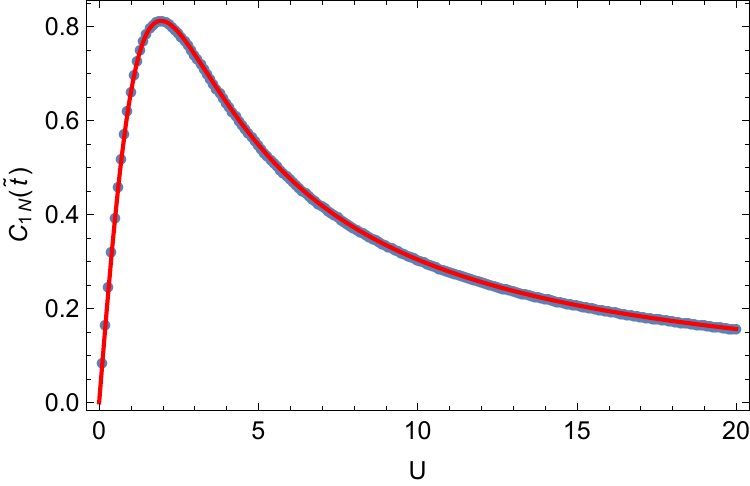}
\caption{Concurrence as a function of $U{=}U^{\uparrow\downarrow}/J$ and the prediction \eqref{concchain}, for a chain of length $N{=}51$. 
The maximal concurrence $f_{1N}^4{=}0.81$ appears for when $U{=}U_{\rm opt}{=}0.97$.}
\label{fig:ConcurrenceFuncU}
\end{figure}

The entanglement generation between the boundaries is evaluated via the 
concurrence \cite{Concurrence} 
$C_{1N}(\tilde{t}){=}2\vert
A^{\uparrow\downarrow}_{1N}(\tilde{t}) A^{\uparrow\downarrow}_{N1}{}^*(\tilde{t})\vert$. 
From the asymptotic analysis \cite{article:Banchi2011}, since the wavepackets are peaked 
around $p_j{\simeq}{\pm}\pi/2$, we find that 
\begin{equation}
C_{1N} {=} f_{1N}^4 \frac{2 U/U_{\rm opt}}{(U/U_{\rm opt})^2+1}~,
\label{concchain}
\end{equation}
where $f_{1N}$ is the transmission probability from site $1$ to site $N$ at the
transmission time and $1{-}U_{\rm opt}{\propto} \Delta^2$, where $\Delta$ is the width of the wave-packet. 
For optimal values \cite{article:Banchi2011} of $J_0{\approx}1.03 N^{-1/6}$ one finds 
$\Delta{\simeq}0.53 N^{-1/3}$ and accordingly 
$U_{\rm opt}{\approx}1{-}0.41 N^{-2/3}$, as shown in the inset of 
Fig.~\ref{fig:UoptL}. 
On the other hand, the maximum value of the concurrence, shown in Fig.~\ref{fig:UoptL},
depends only on the
transfer quality $f_{1N}$, which is different from zero even in the thermodynamic limit 
\cite{article:Banchi2011} $f_{1N}{\gtrsim}0.847$ for any $N$. Therefore, in the thermodynamic limit 
the maximal concurrence is $C{=}f^4_{1\infty}{\approx}0.5144$. 
Explicit results for the dependence of the concurrence upon the interaction $U{\equiv}U^{\uparrow\downarrow}/J$ of a finite chain are shown in 
Fig.~\ref{fig:ConcurrenceFuncU}. 

\begin{figure}[t]
\centering
\includegraphics[width=0.95\linewidth]{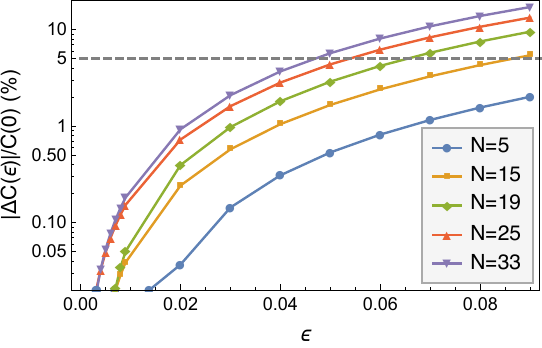}
\caption{Relative variation $| \Delta C(\epsilon) | / C(0)$ of the concurrence between sites (1,$N$) at transfer time, under random diagonal noise with strength $\epsilon$. Several chain lengths $N$ are considered. The gray dashed line represents a threshold of the 5\%.}
\label{fig:RandomNoise}
\end{figure}
Finally, in Fig.~\ref{fig:RandomNoise} we consider the effect of  
noise, in the form of static random local energy shifts in different sites 
added to the Hamiltonian \eqref{e.BH}, 
\begin{equation}
H_{\text{noise}}=\sum_{j,\alpha} \mu_j n_{j,\alpha}~,
\end{equation}
where $\mu_j{=}J x_j$,with $x_j{\in}\left[-\epsilon,\epsilon \right]$ 
is a uniform random distribution and $\epsilon$ is the perturbation strength. We compute  the relative variation of the concurrence with respect to the zero noise case, namely $\vert \Delta C(\epsilon)\vert /C(0)$, where $\Delta C(\epsilon)\equiv C_{1N}^{\rm max}(\tilde{t},\epsilon)-C_{1N}^{\rm max}(\tilde{t},0)$. As it is clear from the figure,  our mechanism is robust against imperfections of $\epsilon \lesssim 0.05$ for a $L=33$ chain.

\section{Concluding remarks}
We have proposed a low control method to generate quantum gates from collision, 
which are necessary building blocks for neutral atom based quantum 
computation. In view of the recent unprecedented capabilities of 
observing atomic quantum walks in lattice experiments 
\cite{GreinerNew,fukuhara}, we show how to use the {\it natural} interaction between atoms
for quantum logic. Our scheme is stable against imperfections and enables the 
realization of 
quantum gates by minimizing the need of external control sources. 
In optical lattice scenarios,
our scheme is compelling for applications,  as the lattice depth
control makes possible to interchange static to flying qubits, avoiding the
necessity to seek some mechanism to couple static to mobile particles. 
At the root of our proposal there is the exploitation of Bethe-Ansatz techniques and 
quantum indistinguishability. 
Compared to other recent proposals for quantum logic in 1D \cite{LahiniNew,Enrico,Enrico2}, our
method is more scalable, as it can use the machinery of integrable models, such
as the Yang-Baxter relation, 
to realize composite operations between multiple particles (see also \cite{Zhang,Zhang2}). 
Indeed, since in integrable models all complex $n$-body
scattering effects can be factorized into two-body $S$ matrices, one can
straightforwardly apply our findings also in multi-particle scenarios.

 {\it Acknowledgements}:--
VK acknowledges financial support by NSF Grant No. DMS-1205422.
SB, LB and EC 
acknowledge financial support 
from the European Research Council under the European Union's Seventh Framework Programme (FP/2007-2013) / ERC Grant Agreement No. 308253.

\newcommand{\hdoi}[2]{\href{https://doi.org/#1}{#2}}


\begin{thebibliography}{99}
\bibitem{comment} Even outside this regime, the gates are still entangling and thereby useful unless we are at the free ($p_A+p_B >> c$) or impenetrable ($p_A+p_B<<c$) limits.
\end{thebibliography}
\end{document}